# Quenching effect of oscillating potential on anisotropic resonant transmission through a phosphorene electrostatic barrier

R. Biswas[a,*] and C. Sinha[a,b]

[a]Department of Physics, P.K.College, Contai, Purba Medinipur, W. B.- 721401, India.

[b]Indian Association for the Cultivation of Science, Jadavpur, Kolkata-700032, India.

**Abstract:**

The anisotropy in resonant tunneling transport through an electrostatic barrier in monolayer black phosphorus either in presence or in absence of an oscillating potential is studied. Non-perturbative Floquet theory is applied to solve the time dependent problem and the results obtained are discussed thoroughly. The resonance spectra in field free transmission are Lorentzian in nature although the width of the resonance for the barrier along the zigzag (Γ-Y) direction is too thinner than that for the armchair (Γ-X) one. Resonant transmission is suppressed for both the cases by the application of oscillating potential that produces small oscillations in the transmission around the resonant energy particularly at low frequency range. Sharp asymmetric Fano resonances are noted in the transmission spectrum along the armchair direction while a distinct line shape resonance is noted for the zigzag direction at higher frequency of the oscillating potential. Even after the angular average, the conductance along the Γ-X direction retains the characteristic Fano features that could be observed experimentally. The present results are supposed to suggest that the phosphorene electrostatic barrier could be used successfully as switching devices and nano detectors.

Key Words:   Phosphorene, Electrostatic barrier, Oscillating potential, Conductance.

[*]Correspondence and request for materials should be addressed to R.B. (email: rbiswas.pkc@gmail.com)

## Introduction:

The ubiquity of resonances in the quantum transport phenomena cannot be overemphasized. The quantum resonant tunneling processes through single or multiple potential barriers using different recently discovered two dimensional (2D) materials [1-3] have drawn significant attention in fundamental research as well as in the field of nano science and high speed electronics. For a successful design of devices based on photonics and optical nano technology demands a deep understanding in the resonance phenomena of different kinds. Specifically, the physics of Fano resonance [4] plays the key role in photonics and many other optical phenomena [5]. The Fano resonance (FR) occurs when a discrete quantum state interferes with a continuum band of states and is manifested as an asymmetric line shape profile in the transmission spectrum indicating a sharp transition from the total transmission to reflection. This attractive feature plays a significant role for many switching and sensor devices in the field of opto-electronics and photonics [6]. One of the most technologically potential candidates as a tunneling material is the monolayer phosphorene (usually referred as black phosphorus) mainly because of its unique electronic and opto electronic properties [7-8], in particular, the most basic striking feature being its high on/off ratio and high carrier mobility at room temperature and atmospheric pressure, the most desirable property for the progress of nano science. The electronic structure of phosphorene guided by weak Van-der-Waals interaction exhibits an anisotropic puckered honeycomb structure (due to $sp^3$ hybridization) having a finite layer dependent band gap originating due to relatively strong spin orbit interaction [9-11]. This finite band gap of phosphorene has made it particularly suitable for different switching devices, unlike the graphene. The structural anisotropy [12-15] along the armchair and zigzag directions of phosphorene, e.g., the puckered arrangement in the former, where the low energy massive quantum particles obey Dirac like linear spectrum, while, a bilayer structure in the latter obeying Schrödinger like quadratic spectrum, has made the material much more favorable than many other recently observed 2D materials for the next generation electronic and opto electronic world. In particular, the aforesaid unique anisotropic electronic structure makes phosphorene very attractive for fundamental research as well as for many technological applications [16], e.g., transistor, detectors and sensors etc.

The present work mainly addresses the ballistic resonant tunneling of electrons through quasi-bound state of phosphorene n-p-n junction (barrier) particularly at below barrier electron energies. Main emphasis is given on the study of anisotropy in the tunneling conductance with respect to the zigzag and the armchair direction both in the absence and presence of an oscillating potential. As is well known, the presence of the quasi-bound (QB) states plays the key role for the resonant tunneling phenomena [17] and the width and the lifetime of the QB states should be intimately related to the width of the resonance and tunneling time through the barrier. The main focus of the present study is the appearance of sharp asymmetric Fano resonance in presence of the oscillating barrier for the low barrier height (BH) and low barrier width (BW). While in contrast, in case of simple n-p-n barrier, a Breit Wigner type resonance (with much broader peak) is noted for higher BW and higher BH due to the presence of the QB states.

The oscillating scalar barrier is treated theoretically in the framework of Floquet theory where the side bands inherent in the Floquet expansion contribute to different inelastic scattering corresponding to absorption and emission of photons while the central band contributes solely the elastic process in the tunneling phenomena [18]. The occurrence of the asymmetric Fano resonance peak for lower BW and BH could be attributed to the coupling between the continuum state of the incident electron and the QB state inside the barrier. When the energy of the incident electron matches with one of the positive energy QB state having a finite width inside the barrier, the Fano resonance appear, the peak width of the resonance being govern by the width of the QB state.

In view of all these, such tunneling devices comprising a gated monolayer phosphorene (MP) and a time dependent perturbation (oscillating potential) are also expected to have a high potentiality in the field of tunable photo detectors [19], apart from aforesaid many other practical importance.

A wealth of vast literature exists [20-30] in experimental and especially in theoretical research on transport phenomena of a monolayer phosphorene or phosphorene nano-ribbons using various kinds of scalar and vector barriers. Regarding specific theoretical works along similar line with respect to the present one, a few studies exist in the literature [22, 23, 28] studying with the ballistic transport properties of monolayer phosphorene in variety of backgrounds. In what follows only a few recent allied works dealing with the anisotropic properties. Anisotropy in transport along the zigzag and armchair direction was studied in MP with single and multiple barriers using both continuum and lattice model. Anisotropic tunneling resistance was also studied in MP under a magnetic barrier, finding that the conductance of the system strongly depends on the orientation of the magnetic barrier and suffers maximum suppression for the magnetic barrier parallel to the armchair direction. Tunneling anisotropy with respect to the armchair and zigzag direction in MP were also studied for different types of potential barriers. In another work [30] Fano resonances (anti resonances) were noted due to the coupling between the vacancy induced localized state and the continuum state of the edges in zigzag phosphorene nano ribbons.

In a very recent work the present authors [29] studied the laser driven (polarized along the zigzag direction) transmission properties through a vector barrier. The laser assisted conductance along the armchair direction exhibits remarkable changes depending on the width of the barrier.

**Theory:**

In the present model, we consider the electron transport through a phosphorene based oscillating electrostatic barrier (Figs. 1(a-c)) developed along the x-direction (armchair direction) of the phosphorene plane (x-y plane). The electrostatic barrier can be realized experimentally by mounting a top gate with constant electric potential ($V_0$) on the 2D puckered surface, where the gate voltage $V_0$ shifts the energy bands (conduction and valence bands) in phosphorene [16]. The barrier height and width can be modulated respectively by controlling the strength of the gate potential and the linear dimension of the gate plate. The time periodic (oscillating) modulation of the barrier height can be achieved from a small AC signal v(t) applied simultaneously to the gate [31] electrode.

Tunneling transport through a phosphorene based electrostatic barrier is governed by the low energy Hamiltonian [9, 29, 32-35] near the Γ point (centre of the Brillouin zone) as,

$$H(t) = \begin{pmatrix} E_c + \eta_c k_x^2 + \nu_c k_y^2 + V(x,y,t) & \gamma k_x \\ \gamma k_x & E_v - \eta_v k_x^2 - \nu_v k_y^2 + V(x,y,t) \end{pmatrix} \quad (1)$$

where the conduction ($E_c$) and the valence ($E_v$) band edges lead to the direct band gap energy $E_g = E_c - E_v$ (~2 eV [33]). The strength of the coupling between the electron and the hole bands are represented by the symbol γ. On the other hand, the terms containing anisotropic band effective masses are represented by $\eta_c, \eta_v, \nu_c$ and $\nu_v$. To preserve the time reversal symmetry, the off diagonal element of the Hamiltonian contains the γ term only [32, 35]. On account of the buckling along the armchair direction (the x-axis), the Hamiltonian contains both $k_x$ and $k_x^2$ terms, in sharp contrast to the presence of only the quadratic term of momentum along the zigzag direction ($k_y$). As a result, the band structure of phosphorene is highly anisotropic along the armchair and zigzag directions. It should be mentioned here that we have neglected the spin orbit coupling (SOC) term in the Hamiltonian equation (1). This is quite justified as the spin orbit coupling strength for phosphorene, consisting of light atoms is very low ~ 1meV which is much lower than the large band gap of phosphorene (~ eV). Further, as for the extrinsic Rashba coupling, the SOC was reported to be ~ μeV, close to the Γ point for the strength of the electric field ~ 1V/nm [29]. The potential profile for the single barrier in case of tunneling along the Γ-X direction (the potential profile being uniform along the y-direction), can be considered (Fig. 1a)s as

$$V(x, y, t) = V(x, t) = V_0 + V_t Cos\omega t \quad \text{for } 0 \leq x \leq d; \quad \text{Region – II} \qquad (2)$$
$$= 0 \qquad \text{elsewhere;} \qquad \text{Region – I and Region – III.}$$

where $V_0$ being the height of the static electrostatic barrier, $V_t$ being the amplitude of the oscillating potential of frequency $\omega$.

Let us first concentrate on the transmission along the armchair direction (the x-direction). For the present scattering problem (both elastic and inelastic channels) we have to solve the time dependent Schrödinger equation $H(t)\Psi^{II}(x, y, t) = i\hbar \frac{\partial \Psi^{II}(x,y,t)}{\partial t}$, with $H(t)$ given by equations (1) and (2). Since the Hamiltonian $H(t)$ is periodic in time we can apply the Floquet theorem [18, 36] to write the wave function as $\Psi^{II}(x, y, t) = \chi(x, y, t) \exp(-iE_F t/\hbar)$, $E_F$ being the Floquet energy and $\chi$ being a periodic function of time such that $\chi(x, y, t + T) = \chi(x, y, t)$, with $T (= 2\pi/\omega)$ as the time period of the oscillating potential.

Following our earlier work [37, 38] the function $\chi(x, y, t)$ can be written as

$$\chi(x, y, t) = e^{-i(E_X - E_F)t/\hbar} e^{-i\alpha Sin(\omega t)} \varphi(x, y) \qquad (3)$$

with $\alpha = V_t/\hbar\omega$. The time periodicity of $\chi$ leads to the parameter $E_X = E_F + m\hbar\omega$, $m$ being an integer. Here $\varphi(x, y)$ is the two component spinor part representing the spatial part of the total wave function. Since the Hamiltonian is invariant along the y-direction (the electrostatic potential being uniform along the y-direction) the momentum component $k_y$ is conserved and the corresponding solution is given by $e^{ik_y y}$. In view of this, the full space part can be written as $\varphi(x, y) = [\varphi_a(x), \varphi_b(x)]^T e^{ik_y y}$, where the x-component of the spinor parts $\varphi_a(x)$ and $\varphi_b(x)$ satisfy the following coupled differential equations

$$\left[\eta_c \frac{\partial^2}{\partial x^2} + \rho_c^m\right] \varphi_a^m(x) = -i\gamma \frac{\partial \varphi_b^m(x)}{\partial x} \qquad 4(a)$$

$$\left[\eta_v \frac{\partial^2}{\partial x^2} - \rho_v^m\right] \varphi_b^m(x) = i\gamma \frac{\partial \varphi_a^m(x)}{\partial x} \qquad 4(b)$$

with $\rho_c^m = (E_F + m\omega - V_0 - E_c - v_c k_y^2)$ and $\rho_v^m = (E_F + m\omega - V_0 - E_v - v_v k_y^2)$. Here the superscript '$m$' used for the spinors $\varphi_{a,b}^m$ refers to the Floquet side band index. Decoupling the above pair of equations leads to a fourth order differential equation for the component $\varphi_a^m(x)$ as

$$\left[a_1 \frac{\partial^4}{\partial x^4} + b_1 \frac{\partial^2}{\partial x^2} + c_1\right] \varphi_a^m(x) = 0 \qquad (5a)$$

where $a_1 = \eta_c \eta_v$, $b_1 = \eta_v \rho_c^m - \eta_c \rho_v^m - \gamma^2$ and $c_1 = -\rho_c^m \rho_v^m$.

Similarly, the other component $\varphi_b^m(x)$ can be obtained from the differential equation

$$\left[a_2 \frac{\partial^3}{\partial x^3} + b_2 \frac{\partial}{\partial x}\right] \varphi_a^m(x) = c_2 \varphi_b^m(x) \qquad (5b)$$

where $a_2 = i\gamma a_1$, $b_2 = i\gamma(\eta_c \rho_v^m - \gamma^2)$ and $c_2 = \rho_v^m \gamma^2$.

The general solutions of eqns. (5a) and (5b) give the spinor function as

$$\begin{pmatrix}\varphi_a^m(x) \\ \varphi_b^m(x)\end{pmatrix} = A_{2m} \begin{pmatrix}1 \\ \xi_1^m\end{pmatrix} e^{k_1^m x} + B_{2m} \begin{pmatrix}1 \\ \xi_2^m\end{pmatrix} e^{k_2^m x} + C_{2m} \begin{pmatrix}1 \\ \xi_3^m\end{pmatrix} e^{k_3^m x} + D_{2m} \begin{pmatrix}1 \\ \xi_4^m\end{pmatrix} e^{k_4^m x} \qquad (6)$$

Here $A_{2m}, B_{2m}, C_{2m}$ and $D_{2m}$ are the constant coefficients corresponding to $m$-th side band in region II; $k_i^m$'s are the solutions for the qarktic equation [39]

$$a_1(k_i^m)^4 + b_1(k_i^m)^2 + c_1 = 0$$

and $\xi_i^m = \frac{1}{c_2}[a_2(k_i^m)^3 + b_2 k_i^m]$, for $i = 1, 2, 3$ and $4$.

Finally, the full solution for the time dependent wave equation corresponding to the Hamiltonian eqn. (1) in the barrier (Region-II) takes the form

$$\Psi^{II}(x, y, t) = \sum_{m,n} \left[ A_{2m} \begin{pmatrix}1 \\ \xi_1^m\end{pmatrix} e^{k_1^m x} + B_{2m} \begin{pmatrix}1 \\ \xi_2^m\end{pmatrix} e^{k_2^m x} + C_{2m} \begin{pmatrix}1 \\ \xi_3^m\end{pmatrix} e^{k_3^m x} \right.$$
$$\left. + D_{2m} \begin{pmatrix}1 \\ \xi_4^m\end{pmatrix} e^{k_4^m x} \right] e^{ik_y y} e^{-i(E+n\omega)t} J_{n-m}(\alpha) \qquad (7)$$

where $J_{n-m}$ is the Bessel function of order $(n-m)$ and $E_F$ is replaced by the energy $E$ of the particle incident on the barrier.

On the other hand, the solutions for the field free regions (Region – I and Region – II) are quite simple. Since an electron of energy $E$ absorbs (emits) $n$-quanta of radiation from (to) the oscillating potential, it is scattered (reflected and transmitted) from the barrier with different Floquet side band energies $(E \pm m\omega)$. It is straightforward to write the wave function of the electron in region I and region III as follows.

$$\Psi^I(x,y,t) = \sum_m \left\{ \delta_{m0} A_{1m} \begin{pmatrix} 1 \\ \lambda_1^m \end{pmatrix} e^{q_1^m x} + B_{1m} \begin{pmatrix} 1 \\ \lambda_2^m \end{pmatrix} e^{q_2^m x} + C_{1m} \begin{pmatrix} 1 \\ \lambda_3^m \end{pmatrix} e^{q_3^m x} \right\} e^{ik_y y} e^{-i(E+m\omega)t}$$

$$\Psi^{III}(x,y,t) = \sum_m \left\{ A_{3m} \begin{pmatrix} 1 \\ \lambda_1^m \end{pmatrix} e^{q_1^m x} + D_{3m} \begin{pmatrix} 1 \\ \lambda_4^m \end{pmatrix} e^{q_4^m x} \right\} e^{ik_y y} e^{-i(E+m\omega)t} \quad (8)$$

Here $A_{1m}, B_{1m}$ and $C_{1m}$ are the constant coefficients corresponding to $m$-th side band in region I, whereas $A_{3m}$ and $D_{3m}$ are the same for region III. Further, $q_i^m$'s follow $k_i^m$'s and $\lambda_i^m$'s follow $\xi_i^m$'s with $V_0 = 0$ in eqns. (4a) and (4b). It is should be mentioned here that for the physically acceptable solutions we have considered on the evanescently decaying solutions for regions I and III.

With the complete knowledge of solutions for the three regions one can calculate the transmission coefficient for the $m^{th}$ Floquet side band by applying the continuity of $\Psi(x,y,t)$ and $\hat{v}_x \Psi(\vec{r},t)$ at the two interfaces, where the velocity operator $\hat{v}_x$ is given by

$$\hat{v}_x = \frac{\partial H}{\partial k_x} = \begin{pmatrix} 2\eta_c k_x & \gamma \\ \gamma & -2\eta_v k_x \end{pmatrix}.$$

The expression for the probability current density along the x-direction consistent to the Hamiltonian eqn. (1) is given by

$$J_x = i\left[ \frac{\partial \psi^\dagger}{\partial x} M \psi - \psi^\dagger M \frac{\partial \psi}{\partial x} - i\gamma \psi^\dagger \sigma_x \psi \right] \quad (9)$$

with the matrix $M = \begin{pmatrix} \eta_c & 0 \\ 0 & -\eta_v \end{pmatrix}$. In view of Eqn. (9), one can now calculate the transmission coefficient (given by the ratio of the transmitted current density to the incident one) for the $m^{th}$ side band $(T_m)$ and the total transmission $(T)$ respectively given by

$$T_m = \frac{\{2p_1^m(\eta_c - \eta_v \lambda_1^m \lambda_1^{m*}) + \gamma(\lambda_1^m + \lambda_1^{m*})\}}{\{2p_1^0(\eta_c - \eta_v \lambda_1^0 \lambda_1^{0*}) + \gamma(\lambda_1^0 + \lambda_1^{0*})\}} \left| \frac{A_{3m}}{A_{10}} \right|^2 \quad \text{and} \quad T = \sum_{m=-\infty}^{m=\infty} T_m \quad (10)$$

where $A_{10}$ and $A_{3m}$ are the amplitudes of the incident and transmitted (for $m^{th}$ side band) wave respectively, $p_1^0$ being the incident wave vector and $q_1^m = ip_1^m$.

Finally, the integration over $k_y$ of the total transmission ($T$, in Eqn.(10)) reproduces the zero temperature conductance ($G$) [29, 40] through the oscillating electrostatic barrier given by

$$G = G_0 \int_{-k_{max}}^{k_{max}} T \frac{dk_y}{2\pi} \quad (11)$$

$k_{max}$ being the maximum value of the y – component of momentum and $G_0$ is a constant having the dimension of conductance.

A similar set of equations may be obtained when the barrier is created along the zigzag direction or the y-directions (i.e. potential is uniform along the x-direction) and are given in the appendix section.

Results and Discussion:

The main motivation of the present work is to study the effect of time dependent oscillating potential on the resonant transmission of electron through phosphorene electrostatic barrier (n-p-n junction), particularly in the below barrier regime. The parameters in all the figures are in dimensionless unit, e.g., energy ($E$) in $E_0 (= 1$ eV), length in $L_0 = \sqrt{\frac{\hbar^2}{2m_0 E_0}}$, frequency of the oscillating potential in $\omega_0 = \frac{E_0}{\hbar}$ and potentials $V_0$ and $V_t$ are in $E_0$. Typically, for a particilar value $E_0 = 1$ eV, the other parameters will be the following; $L_0 = 1.94346 \times 10^{-10}$ m, $k_0 = 5.14547$ m$^{-1}$ and $\omega_0 = 15.25884 \times 10^{14}$ Hz. Before presenting the actual results for the time dependent barrier, let us first discuss in brief, the nature of the field free (FF) transmission of the electron, displayed in Figs. 2(a) to 2(c). The bottom of the conduction band is chosen as the reference level for measuring the energy ($E$). Fig.2(a) depicts the direction dependence of transmission coefficient ($T_c$) for a set of different incident energies $E$ (in units of eV) = 0.9, 1.1, 1.2, 1.5 and 2.0 for $D = 5$ (barrier width in units of nm) and $V_0 = 1$ (barrier height). For all energies, $T_c$ is symmetric with respective to the +ve and −ve glancing incidence and after a limiting value of $k_y$ (say, $k_y^{max}$), depending on $E$, the transmission disappears completely. The limiting value could be attributed to the absence of any propagating mode in the incident channel governed by the energy conservation relation. For low energy ($E \lesssim 1$), the tunneling probability is almost negligible even over the allowed range of $k_y$, probably due to the fact that the incident energy lies within the band bap of region II (Fig.1(c)) and all the component waves are evanescent in nature. However, for higher incident energies ($E > 1.1$), $T_c$ becomes oscillatory in nature with respect to the angle of incidence and exhibits Fabry Perot resonances.

Regarding below barrier transmission, interesting resonant tunneling features occur in the energy dependent transmission spectra when the incident energy lies within the valence band inside the barrier region (Fig.1(b)), i.e., when the barrier height is high. Fig. 2(b) shows the transmission coefficient for normal incidence for $D = 1$ and $V_0 = 1, 2, 3, 4$ and 5. It should be mention here that the quasi-particle in phosphorene is described by a linear combination of four possible solutions [29, 38]. For lower barrier heights (e.g., $V_0 = 1$ and 2), the below barrier incident energy ($0 < E < V_0$) lies within the band gap in region – II, leading to possible four evanescent solutions. Thus $T_c$ increases exponentially with increasing $E$. However, for higher barrier heights (BH) $V_0 = 3$ and 4 corresponding to $E = 0.45$ and 1.363 respectively, the Breit Wigner resonances (BWR) are noted in the transmission spectrum. In fact, for $V_0 > 2$ two possible cases may arise. For the first case, for $0 < E < |E_v|$ both the oscillatory and the evanescent modes coexist in region II and the coherent superposition (depending on the incident energy, barrier height and width) of the two oscillatory modes may lead to the resonance transmission of BWR type. While in the second case $|E_v| < E < E_c$, since only the evanescent solutions exist, no resonant tunneling peak (RTP) arises. It is well known that the presence of the bound or quasi-bound state plays a major role in the transmission through a tunneling structure. As for the nature of the bound state, it may be mentioned that the bound state for E<0 in region II (vide Fig. 1 (c)) is discrete in nature while the state at E = $E_1$ in Fig. 1(b) is of quasi-bound nature with finite lifetime. This is particularly because in the former case the state is bounded by band gap in regions I and III where all the solutions are evanescent in nature, while for the latter (Fig.1(b)), the bounding regions are valence bands that allow propagating modes in the corresponding regions (I and III). Thus the bound state in Fig. 1(b) is termed as quasi-bound state with finite width while for the Fig. 1(c) it is of localized nature, i.e a discrete bound state (DBS). Further, a propagating electron with E> 0 (e.g., E = $E_1$) in region I can resonantly tunnel through the QBS in Fig. 1(b). Hence they are also termed as tunneling quasi-bound state. In contrast, the bound state in Fig. 1 (c) has no effect on the resonant transmission under the field free condition, although it plays a major role in photon assisted transmission to be discussed later. With decreasing $V_0$, the position of the RTP moves towards the lower energy end of the spectrum and finally disappears. Since the position of the RTP corresponds to the position of the quasi-bound state, we may conclude that with the decrease in BH the quasi-bound state shifts below the reference energy level resulting in the disappearance of the RTP at lower value of $V_0$. The presence of RTP indicates that in phosphorene electrostatic barrier, the electron can resonantly tunnel through the hole like quasi-bound state lying within the valence band inside the barrier (region-II).

Regarding the effect of barrier width (BW), no RTP is noted for $D \lesssim 1$, corresponding to the parameter $k_y=0$ and $V_0= 3$ (vide Fig. 2(c)). However, for higher values of $D$, e.g., 1.5 and 2.0 one RTP is noted in each case. With the increase in the width of the barrier the RTP shifts towards the higher energy end of the transmission spectrum. This is quite legitimate since in order to maintain the phase coherence under wider barrier condition, one has to increase the wavelength of the particle which in turn reduces its energy and thereby moves towards the bottom of the valence band ($E_v$) in region II. Further, gradual decrease in width of the RTP indicates that the valence band quasi-bound state gets thinner with the increase in $D$.

With a view to study the anisotropy, we present in Figs 3(a) and 3(b) the field free (FF) $T_c$ for the barrier along the Γ-Y direction. The angular range for transmission along the zigzag direction (y) is shrunk compared to those along the armchair direction (x) [22]. In the present case (zigzag) the angular transmission profile is more oscillatory in nature and the number of resonances is greater for tunneling along the y-direction than along the x-direction. Further, the energy dependent transmission (vide Figs. 3(b)) reveals that the above barrier transmission for the zigzag case is highly oscillatory and for below barrier energies, $T_c$ is vanishingly small except at resonances (below barrier resonances are noted for $V_0 > 2$). Regarding the resonances, one remarkable difference in $T_c$ noted between the armchair and the zigzag direction is that the width of the RTP (for the below barrier case) is much smaller for the latter than that for the former one. Thus for the barrier along Γ-Y, the quasi-bound states in the valance band inside the barrier might be of narrow width.

The primary concern of the present work is to study the effect of the time dependent oscillating potential on the resonant tunneling transport of electron through a phosphorene electrostatic barrier. With a view to this, we display in Fig. 4(a), the total transmission probability $T$ $(= \sum T_m)$ through the barrier corresponding to the tunneling parameters $D = 4$, $k_y= 0$ and $V_0 = 2.5$ for which we have only one RTP at $E \sim 0.3$ in absence of the oscillating field. The frequency of the applied field is chosen as low as $\omega = 0.003$ lying in the Tz region (frequency is expressed in units of $\omega_0 = 1.5 \times 10^{15}$ Hz). For the oscillating potential (OP) of lower amplitude, e.g., $V_t = 0.01$ (or 10 meV), the height of the RTP reduces from that for the FF while the position and shape remain almost unchanged. With further increase in $V_t$, the shape of $T_c$ deviates from the BWR type with the appearance of additional two small ridges almost placed symmetrically on either side of the FF peak. The amount of displacement (for $V_t = 0.03$) indicates that the new peaks could probably arise due to the absorption (high energy ridge) and emission (low energy ridge) of single photon. Fig. 4(a) reveals that the FF transmission is systematically suppressed by the application of the oscillating potential. Regarding the frequency variation of $T_c$ (in the Tz range), we note that the quenching of $T_c$ is more pronounced at lower frequency of the oscillating potential (Fig. 4(b)).

A different type of quenching of the resonance is noted at higher values of the frequency (ω) of the oscillating potential. Fig. 5(a) depicts the energy dependent transmission $T$ (total transmission summed over all Floquet side bands) for the parameters $D = 1$, $k_y = 0$, $V_0 = 4$ and ω = 1 for which the below barrier FF transmission exhibits only one RTP (BWR type) corresponding to the QBS at energy $E_b = 0.495$. With application of the OP ($V_t = 1.0$), the magnitude of the RTP reduces from 1.0 (FF) to 0.6 while the shape and the position remains unchanged. Another small peak of height 0.2 is noted at $E \sim 1.5$, the resonance arises due to the absorption of one photon inside the barrier. With further increase in $V_t$, multi-photon induced resonance peaks occur at energies governed by the conservation relation $E = E_b \mp n\omega$, $n$ being the number of photons absorbed/emitted by the charge carrier. Thus with the increase in $V_t$, the probability of the photon induced resonances gradually increases at the expense of the FF resonance. For $V_t = 2.5$, $T$ changes from 1.0 to 0.1, indicating that the FF resonance is almost suppressed by the oscillating field. The occurrence of the resonant peaks due to photon exchange processes indicates the possibility of creation of virtual QBS within the forbidden energy gap inside the barrier. The knowledge of tailoring the QBS by the application of the oscillating potential could be of potential importance in the fabrication of phosphorene based nano devices.

We now discuss (vide Fig. 5(b)) the field effects on $T$ along Γ-X when the FF transmission does not exhibit any resonance, e.g., the case of $D = 1$, $k_y = 0$ and $V_0 = 3$. Although a slow monotonic rise in $T_c$ is noted in the FF results (Fig. 2(b)) within $0 < E < V_0$, the OP induces the exotic Fano resonances having asymmetric line profile (a sharp maximum followed by a minimum) in the transmission spectrum. This asymmetry originates from the close coexistence of resonant reflection and resonant transmission. As is well known [4, 41,42], the FR arises due to the coherent coupling (interference) between a localized (discrete) channel and a continuum channel. In the present oscillating single barrier case, the appearance of the FR (Fig. 5(b)) indicates that there must be a discrete localized state inside the barrier corresponding to energy E~ -0.3 and the photon mediated quantum coherent coupling between the incident continuum and the discrete bound state (Fig.1(c)) results in the appearance of the characteristic FR. In fact, the bound state gets dressed by the absorption or emission of photon to rise energetically close to the incident continuum and interact resonantly with the continuum state. In contrast, the BWR occurs due the perfect transmission of electron through the tunneling QBS (Fig.1(b)) having finite with existing inside the barrier as its energy matches exactly with the incident energy. Since the aforesaid QBS (Fig. 1(b)) inside the barrier is of finite width there is no chance of FR to occur as it does not satisfy the criteria of the FR. Multi-photon induced FR's, although less probable than the single photon one, are also noted for comparatively lower frequency $\omega$, which is quite expected. The appearance of the field induced FR in the tunneling transport along the armchair ($\Gamma$-X) direction indicates the existence of a DBS in contrast to the tunneling QBS of finite width for the FF Breit Wigner resonance. This is quite justified physically because for the BWR case, the QBS inside the barrier (Region-II) is energetically enclosed (Fig. 1(b)) by the conduction band on both sides (regions I and III) while for the FR, both the regions I and III correspond to band gap (Fig. 1(c)).

In order to study the effect of the OP on the electron transport along the zigzag ($\Gamma$-Y) direction, we display in Figs. 6(a) to 6(c) the energy dependent total transmission probability (*T*) particularly for the below barrier condition of the system. Figs. 6(a) and 6(b) emphasize the transmission resonance for different values of the amplitude and frequency of the oscillating potential respectively. Fig. 6(a) reveals that for at low frequency (say, $\omega$~0.003), the height of the FF RTP is reduced gradually (keeping the position almost same) with the increase in the amplitude of OP ($V_t$). However, beyond a certain value of $V_t$, the Lorentzian type resonance disappears completely and is replaced by a simple oscillatory transmission with increasing number of periods with higher $V_t$. Further, the amplitude of oscillation decreases with the decrease in $\omega$, vide Fig. 6(b). Comparing *T* along the $\Gamma$-X and $\Gamma$-Y directions it may be concluded that although the RTP is suppressed in both the cases the latter is more oscillatory than the former near the position of resonance (up to the maximum value of $\alpha$=10 considered here).

As regards the OP effect on the below barrier transmission along the zigzag (Γ-Y) direction, characteristic line spectra resonances are noted (vide Fig. 6(c)) under $D = 1$, $k_x = 0.001$ and $V_0 = 2$ for which the FF spectrum does not exhibit any transmission resonances. The energy nodes ($E = 0.27$, $0.47$ and $0.77$) corresponding to the lines obtained for the frequencies $\omega = 0.5$, $0.7$ and $1.2$ reveal that the FF barrier contains a bound state at energy $E_b = -0.23$ similar to that for the armchair direction. The only difference lies in the appearance of the line type resonance in place of the FR observed for the latter (Γ-X). This difference could probably be attributed to the fact that the amplitude of the evanescent modes of propagation (along the Γ-Y direction) are too small to interfere even in presence of photon.

Since the only measurable quantity for tunneling is the conductance, it is highly desirable to calculate the corresponding conductance in order to study its behavior in presence of the oscillating potential. Figs. 7(a) and 7(b) display the zero temperature conductance (total transmission integrated over all incident angles) corresponding to the oscillating electrostatic barriers in phosphorene oriented along the Γ-X and Γ-Y directions respectively. Regarding armchair (Γ-X) conductance, Fig. 7(a) reveals that the energy dependent field free $G$ (in units of $G_0$) under the below barrier condition is resonant-like in nature. Comparing Figs. 2(b) (green dash line) and 7(a) (black solid line) it reveals that the quantitative nature of the conductance and the transmittance (at normal incidence) is almost similar while quantitatively, the width of the resonance increases and the position of the maximum is red shifted in the former as compared to the latter. This feature is quite justified since under the condition $E < V_0$, the transmission decreases sharply away from the normal incidence. As for the effect of the OP, it may be noted (Fig. 7(a)) that at lower value of $V_t$, the quantitative nature of $G$ remains almost unaffected except for a small decrease in magnitude near the resonant energy only. On the other hand, with the increase in $V_t$, the resonant nature of the conductance profile changes gradually to an oscillatory one with a systematic suppression of conductance through the barrier, particularly around the resonant peak. However, away from resonance, the magnitude of G increases in presence of the oscillating field. In fact with the increase in $V_t$, multiphoton processes become more probable whereby the position of the overall maxima is shifted gradually towards the higher energy end of the conductance profile. We thus conclude that the electron transmission integrated over the incident angle is quenched significantly by the application of OP. This control of the conductance by tuning of the oscillatory field parameters finds a vital importance for the exploitation of phosphorene in switching and sensor devices. Regarding the nature of $G$ along the zigzag ($\Gamma$-Y) direction it may be noted from the Fig. 7(b) that the FF conductance exhibits sharp resonance, unlike the nature of discrete line shape observed at the transmission level. The effect of the OP is to suppress the resonance conductance showing small oscillation around the maximum of the FF conductance.

Apart from the quenching effect at resonance, the specialty of the photon induced scattering through the barrier along the $\Gamma$-X direction is the appearance of the asymmetric FR in the transmission spectra as already mentioned earlier. Fig. 8 displays the below barrier conductance spectra in phosphorene oscillating electrostatic barrier along the $\Gamma$-X direction for the structural parameter $D = 1$ and $V_0 = 3$ for which the time independent barrier exhibits monotonic increase in transmission (Fig. 2(b)) in contrast to the asymmetric FR noted (Fig. 5(b)) for the time dependent one. The conductance profile (Fig. 8) clearly depicts the presence of an asymmetric FR of larger width ($\Delta E \sim 0.473 - 0.277$, for $\omega = 0.6$) unlike the sharp feature noted in the transmission spectra (Fig. 5(b)). However, in most of the earlier works [43] the characteristic FR was poorly reflected at the conductance level in case of bilayer graphene (BLG) because this distinctive signature smears out due to method of the angular integration of the transmittance. This indicates the demand of the measurement at the differential level of the conductance [43]. In the present case, however, peak of the FR in the conductance is pronounced and the position ($E \sim 0.473$) coincides with that for the case of normal incidence (at the transmittance level) as expected, although the (non-zero) conductance minimum of the FR does not correspond to the corresponding (zero) minimum of the transmission profile. This seems to be quite legitimate since with increasing $k_y$ the position of the FR minimum moves towards the lower energy end of the transmission spectrum which on angular average (i.e., in the conductance profile) appears to be smeared out to some extent. Very recently it is reported [44] that the opening of a band gap in BLG can exhibit a Fano profile even in the conductance (integrated over entire angular range) due to improvement of the chirality matching between extended and discrete states [44]. The noted FR in the conductance profile is particularly significant in the sense that it could be possible to observe experimentally. On changing the frequency from 0.6 to 0.8 the FR maximum of the conductance spectra is shifted exactly by the same amount (i.e., $\Delta E = 0.2$). It is thus evident that from the measurement of conductance one can find the unknown frequency and the electrostatic barrier can act as a frequency detector.

Conclusion:

The most salient features of the present study that deals with the tunneling transport across a phosphorene electrostatic barrier both in absence and in presence of a time dependent oscillating potential are as follows. It deals with the behavior of tunneling transmission and conductance with respect to the structural asymmetry of phosphorene along the armchair and zigzag directions. The below barrier field free (FF) transmission exhibits interesting directional asymmetry, e.g., for lower $V_0$, the barrier along Γ-X is more transparent towards charge transport as compared to the Γ-Y barrier. The FF Breit-Wigner resonance is comparatively sharper for the Γ-Y barrier than for the Γ-X case. As regards the effect of the oscillating field, the quenching of the resonant transmission is pronounced for both the cases and the magnitude of suppression increases with the increase in amplitude of the oscillating potential. At low frequency, the FF resonant transport turns out to be oscillatory, the number of oscillations being greater for the Γ-Y barrier than the Γ-X one. The major directional anisotropy in transmission lies in the fact that at high frequency regime, the barrier along the Γ-X direction exhibits sharp asymmetric Fano resonances in sharp contrast to the line spectrum noted for the Γ-Y barrier. Another major finding of the present work is the prediction for the distinct asymmetric FR profile in the zero temperature conductance spectrum in the armchair direction in contrast to the zigzag one. The photon mediated quenching effect noted in the present work at lower frequency of the oscillating potential highlights the possibility of using the phosphorene electrostatic barrier as a switching device. On the other hand, the prediction of FR at the conductance level at higher frequencies, support the exploitation of the structure in the fabrication of phosphorene based nano- sensor and detector. To our knowledge, the measurement of tunneling conductance under the oscillating potential is not yet available in the literature although it might come out in near future using phosphorene or gapped graphene bilayer.

APPENDIX:

For tunneling along the Γ-Y direction, the potential profile being uniform along the x-direction one can write

$$V(x, y, t) = V(y, t) = V_0 + V_t Cos\omega t \quad \text{for } 0 \leq y \leq d; \quad \text{Region – II} \tag{A1}$$
$$= 0 \quad \text{elsewhere;} \quad \text{Region – I and Region – III.}$$

In that case the coupled differential equations for the two pseudo-spin components are given by

$$\left[v_c \frac{\partial^2}{\partial y^2} + \rho_c^m\right] \varphi_a^m(y) = \gamma k_x \varphi_b^m(y) \tag{A2)a}$$

$$\left[v_v \frac{\partial^2}{\partial y^2} - \rho_v^m\right] \varphi_b^m(y) = -\gamma k_x \varphi_a^m(y) \tag{A2)b}$$

with $\rho_c^m = (E_F + m\omega - V_0 - E_c - \eta_c k_x^2)$ and $\rho_v^m = (E_F + m\omega - V_0 - E_v + \eta_v k_x^2)$.

The solutions for the spinor functions take the form of Eqn. (6), where $k_i^m$'s are the solutions for the qarktic equation

$$f_1(k_i^m)^4 + g_1(k_i^m)^2 + h_1 = 0 \tag{A3}$$

with $f_1 = v_c v_v$, $g_1 = \rho_c v_v - v_c \rho_v$ and $h_1 = (\gamma k_x)^2 - \rho_c \rho_v$.

Here $\xi_i^m = \frac{1}{h_2}[f_2(k_i^m)^2 + g_2]$, for $i = 1, 2, 3$ and 4 with $f_2 = f_1$, $g_2 = \rho_c v_v$ and $h_2 = \gamma v_v k_x$.

Further the expression for the probability current density along the x-direction corresponding to the Hamiltonian eqn. (1) is given by

$$J_y = i\left[\frac{\partial \psi^\dagger}{\partial y} N\psi - \psi^\dagger N \frac{\partial \psi}{\partial y}\right] \tag{A4}$$

with the matrix $N = -2 \begin{pmatrix} v_c & 0 \\ 0 & -v_v \end{pmatrix}$.

Finally, the transmission coefficient is given by

$$T_m = \frac{p_1^m \{v_c - v_v \lambda_1^m \lambda_1^{m*} + \beta(\lambda_1^m + \lambda_1^{m*})\}}{p_1^0 \{v_c - v_v \lambda_1^0 \lambda_1^{0*} + \beta(\lambda_1^0 + \lambda_1^{0*})\}} \left|\frac{A_{3m}}{A_{10}}\right|^2 \tag{A5}$$

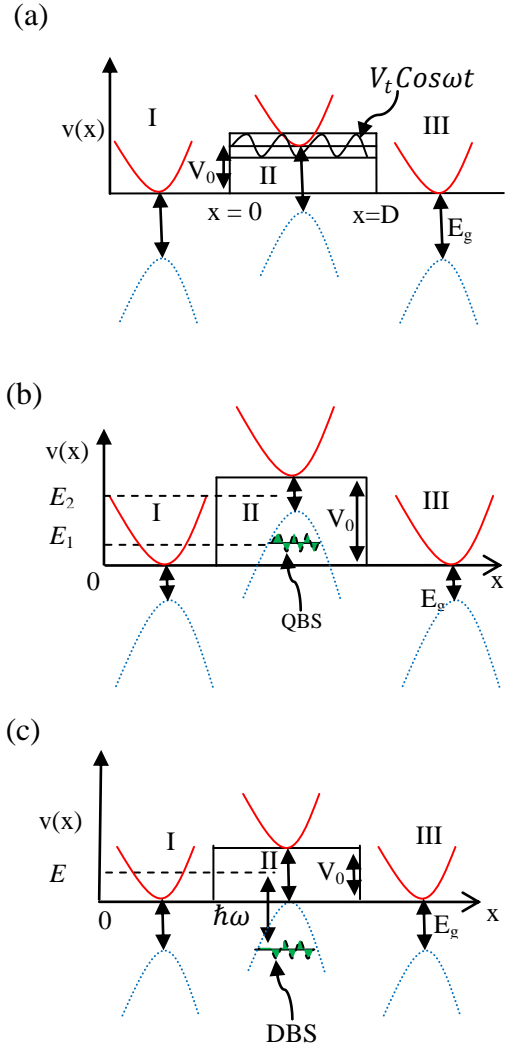

Fig.1: Schematic diagram for the potential profile along the x-axis of a rectangular electrostatic barrier (corresponding to region - II) of height $V_0$ and width $D$ arising due to shifts of the conduction (solid red line) and valance (dotted blue line) bands relative to the regions I and III by the application of a top electrostatic gate in region II on the phosphorene plane. A small ac signal of amplitude $V_t$ and angular frequency $\omega$ is applied to the gate in region II. Bottom of the conduction band in region I is taken as the reference level (zero) for measuring energy. (a) $V_0 < E_g$, $E_g$ corresponds to the band gap energy. (b) $V_0 > E_g$; quasi-bound state (QBS) exists corresponding to energy E, responsible for the below barrier transmission resonance in static case. (c) $V_0 = E_g$; shows the existence of the QBS responsible for the resonances in the photon assisted transmission.

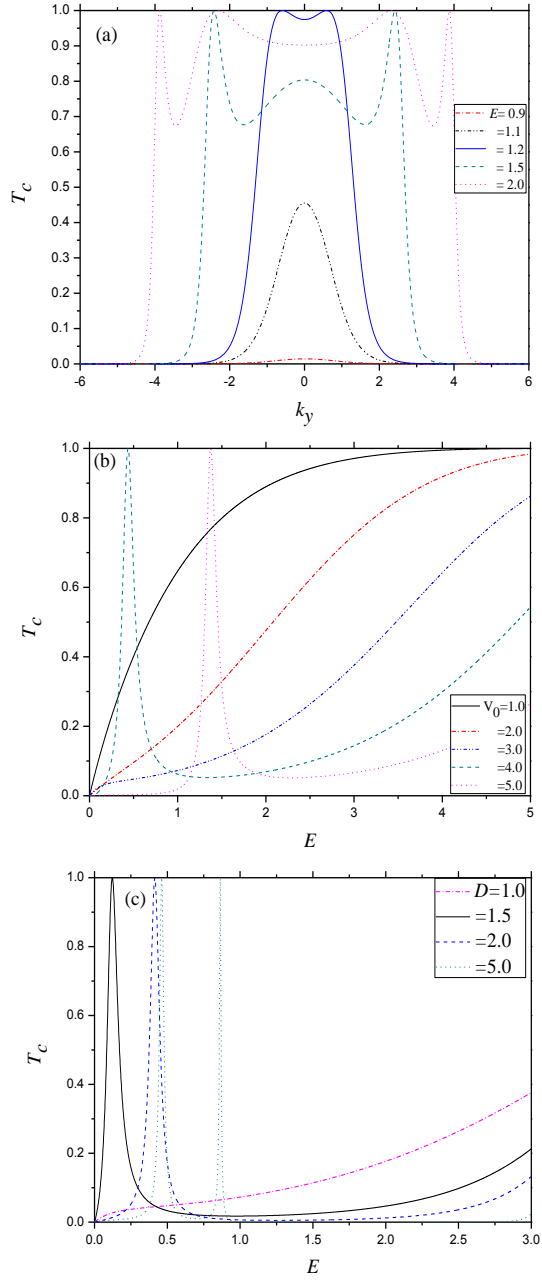

Fig.2: (a) Transmission coefficient ($T_c$) as a function of the y-component of wave vector ($k_y$) through a static phosphorene electrostatic barrier along $\Gamma - X$ direction. For $V_0 = 1$ and $D = 5$. Dash dot (red) line for $E = 0.9$; Dash double dot (black) line for $E = 1.1$; Solid (blue) line for $E = 1.2$; Dash (green) line for $E = 1.5$ and Dot (purple) line for $E = 2.0$. (b) Same as (a) but as a function of incident energy ($E$) for $k_y = 0$ and $D = 1$. Solid (black) line for $V_0 = 1$; Dash dot (red) line for $V_0 = 2$; Dash double dot (blue) line for $V_0 = 3$; Dash (green) line for $V_0 = 4$; Solid (black) line for $V_0 = 1$; Dot (purple) line for $V_0 = 5$. (c) Same as (b) but for $k_y = 0$ and $V_0 = 3$. Dash dot (purple) line for $D = 1$; Solid (black) line for $D = 1.5$; Dash (blue) line for $D = 2.0$; Dot (green) line for $D = 5$.

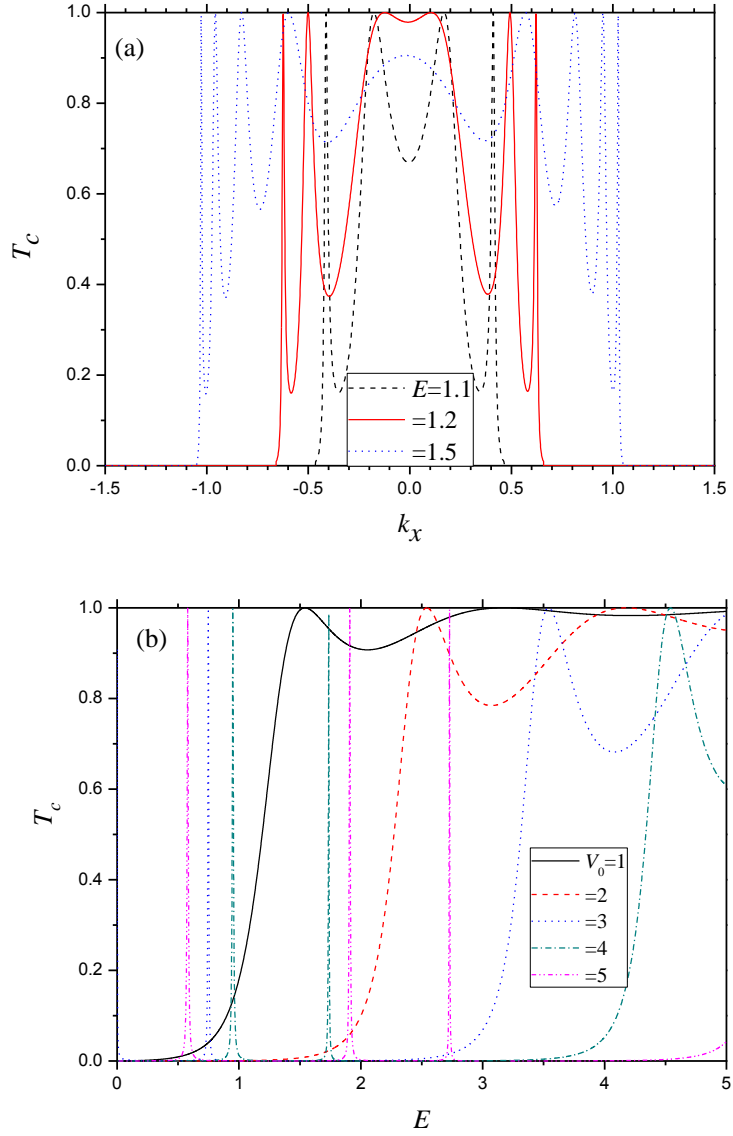

Fig.3: (a) $T_c$ as a function of the x-component of wave vector ($k_x$) through a static phosphorene electrostatic barrier along $\Gamma - Y$ direction for $D = 5$ and $V_0 = 1$. Dash (black) line for $E = 1.1$; Solid (red) line for $E = 1.2$; Dot (blue) line for $E = 1.5$. (b) $T_c$ v.s. $E$ for barrier along $\Gamma - Y$ direction with $D = 1$ and $k_x = 0.001$. Solid (black) line for $V_0 = 1$; Dash (red) line for $V_0 = 2$; Dot (blue) line for $V_0 = 3$; Dash dot (green) line for $V_0 = 4$; Dash dot dot (purple) line for $V_0 = 5$.

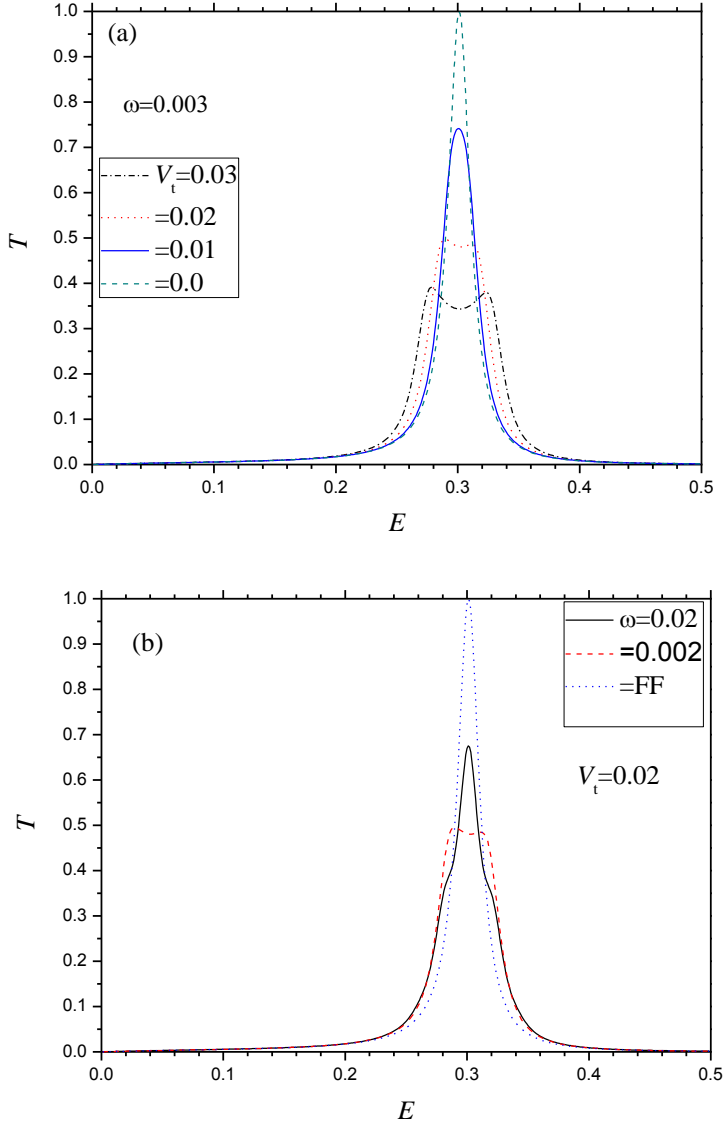

Fig.4: Total transmission $T(=\sum_{m=-\infty}^{m=\infty} T_m)$ v.s. $E$ for an oscillating (frequency in Tz region) electrostatic barrier (e.g. Fig.1(b)) along $\Gamma - X$ direction under below barrier condition for $k_y = 0$, $D = 4$ and $V_0 = 2.5$. (a) For $\omega = 0.03$; Dash (green) line for field free (FF) case (i.e., $V_t = 0$); Solid (blue) line for $V_t = 0.01$; Dot (red) for $V_t = 0.02$; Dash Dot (black) for $V_t = 0.03$. (b) Same as (a) but for $V_t = 0.02$.; Dot (blue) for FF; Dash (red) for $\omega = 0.002$; Solid (black) for $\omega = 0.02$.

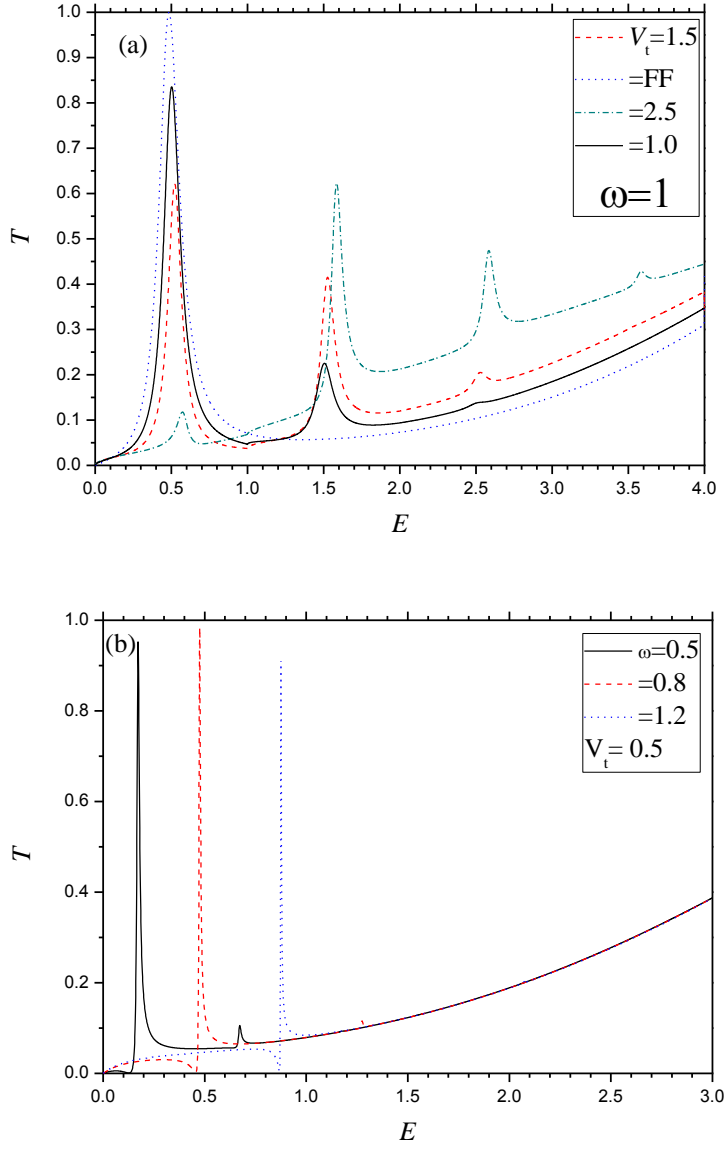

Fig.5: Same as Fig. 4 but for high frequency regime with $k_y = 0$ and $D = 1$. (a) For $V_0 = 4$ and $\omega = 1$. Dot (blue) for FF; Solid (black) for $V_t = 1.0$; Dash (red) for $V_t = 1.5$; Dash dot (green) for $V_t = 2.5$. (b) For $V_0 = 3$ and $V_t = 0.5$. Solid (black) for $\omega = 0.5$; Dash (red) for $\omega = 0.8$; Dot (blue) for $\omega = 1.2$.

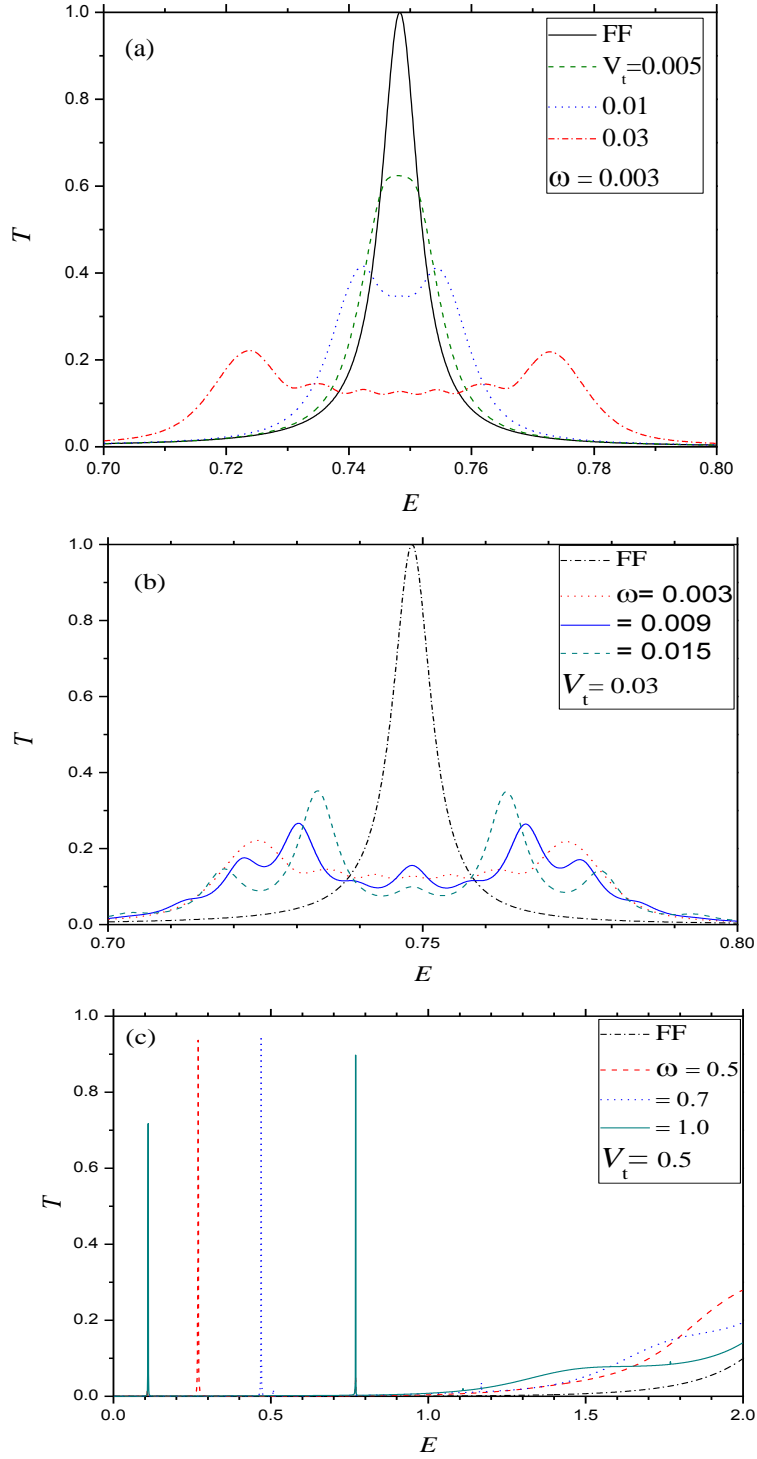

Fig.6: Same as Fig. 4 but for barrier along the Γ − Y direction for $k_x = 0.001$ and $D = 1$. (a) For $\omega = 0.003$ and $V_0 = 3$. Solid (black) for FF; Dash (green) for $V_t = 0.005$; Dot (blue) for $V_t = 0.01$; Dash dot (red) for $V_t = 0.03$; (b) For $V_0 = 3$ and $V_t = 0.03$; Dash dot (black) for FF; Dot (red) for $\omega = 0.003$; Solid (blue) for $\omega = 0.009$; Dash (green) for $\omega = 0.015$. (c) For $V_0 = 2$ and $V_t = 0.5$. Dash dot (black) for FF; Dash (red) for $\omega = 0.5$; Dot (blue) for $\omega = 0.7$; Solid (green) for $\omega = 1$.

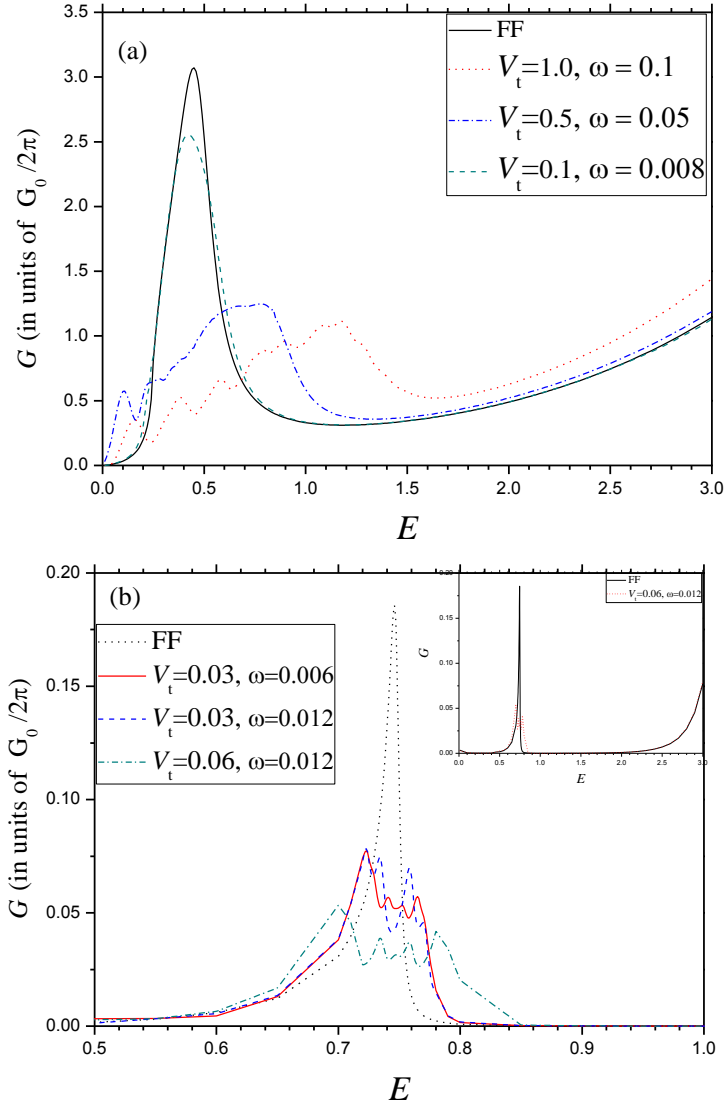

Fig.7: Zero temperature conductance $G$ (in units of $\frac{G_0}{2\pi}$) v.s. $E$ through an oscillating electrostatic in phosphorene. (a) Barrier along the $\Gamma - X$ direction for $D = 1$ and $V_0 = 4$. Solid (black) for FF; Dot (green) for $V_t = 0.1$ and $\omega = 0.008$; Dash dot (blue) for $V_t = 0.5$ and $\omega = 0.05$; Dash (red) for $\omega = 0.1$ and $V_t = 0.1$ (b) Barrier along the $\Gamma - Y$ direction for $D = 1$ and $V_0 = 3$. Dot (black) for FF; Solid (red) for $V_t = 0.03$ and $\omega = 0.006$; Dash (blue) for $V_t = 0.03$ and $\omega = 0.012$; Dash dot (green) for $V_t = 0.06$ and $\omega = 0.012$. INSET: For entire energy range under below barrier condition. Solid (black) for FF and dot (red) for $V_t = 0.06$ and $\omega = 0.012$.

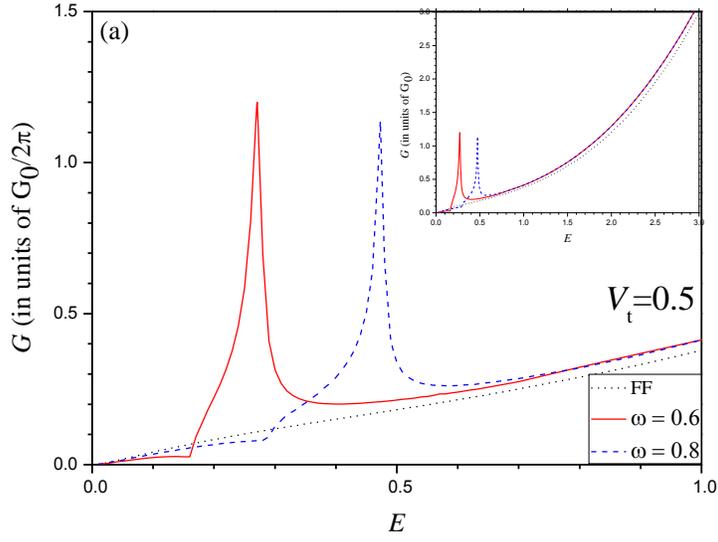

Fig.8: Same as Fig.7 but for the barrier along the $\Gamma - X$ direction for $D = 1$ and $V_0 = 3$. Dot (black) for FF; Solid (red) for $V_t = 0.5$ and $\omega = 0.6$; Dash (blue) for $V_t = 0.5$ and $\omega = 0.8$. INSET: For entire energy range under below barrier condition.